\documentclass[sigconf,nonacm]{acmart}

\AtBeginDocument{%
  }
\usepackage[most]{tcolorbox} 
\usepackage{makecell}
\usepackage{multirow}
\usepackage{xcolor}
\usepackage[linesnumbered,ruled,vlined,noend]{algorithm2e}

\usepackage{amsmath, amssymb}  
\usepackage{booktabs}
\usepackage{graphicx}
\usepackage{threeparttable}

\setcopyright{acmlicensed}
\copyrightyear{2026}
\acmYear{2026}
\acmDOI{XXXXXXX.XXXXXXX}
\acmConference[ICSE '26]{2026 IEEE/ACM 48th International Conference
on Software Engineering}{April 12--18,
  2026}{Rio de Janeiro, Brazil}
\acmISBN{978-1-4503-XXXX-X/2018/06}
\newcommand{\Code}[1]{\begin{small}\texttt{#1}\end{small}}

\tcbset{
  my box/.style={
    enhanced,
    colframe=#1!80,
    colback=#1!10,
    attach boxed title to top left={xshift=0.2cm, yshift=-0.2cm},
    boxed title style={
      colback=#1!80,
      outer arc=0pt,
      arc=0pt,
      top=0pt,
      bottom=0pt,
    },
  },
}
\newtcolorbox{result-rq}[1]{
  my box=black,
  title=#1,
  boxrule=1.2pt,top=6pt,bottom=3.5pt,left=6pt,right=6pt
}

\begin{document}

\title{Small Changes, Big Trouble: Demystifying and Parsing License Variants for Incompatibility Detection in the PyPI Ecosystem}
\author{Weiwei Xu}
\affiliation{
  \institution{
    School of Computer Science, Peking University, and Key Laboratory of High Confidence
    Software Technologies, Ministry of Education}
  \city{Beijing}
  \country{China}
}
\email{xuww@stu.pku.edu.cn}
\author{Hengzhi Ye}
\affiliation{
  \institution{
    School of Computer Science, Peking University, and Key Laboratory of High Confidence
    Software Technologies, Ministry of Education}
  \city{Beijing}
  \country{China}
}
\email{hzye@stu.pku.edu.cn}
\author{Kai Gao}
\affiliation{
  \institution{
    School of Computer \& Communication Engineering, University of Science and Technology Beijing}
  \city{Beijing}
  \country{China}
}
\email{kai.gao@ustb.edu.cn}
\author{Minghui Zhou}
\authornote{Corresponding Author}
\affiliation{
  \institution{
    School of Computer Science, Peking University, and Key Laboratory of High Confidence
    Software Technologies, Ministry of Education}
  \city{Beijing}
  \country{China}
}
\email{zhmh@pku.edu.cn}


\begin{abstract}
Open-source licenses establish the legal foundation for software reuse, yet license variants, including both modified standard licenses and custom-created alternatives, introduce significant compliance complexities. Despite their prevalence and potential impact, these variants are poorly understood in modern software systems, and existing tools do not account for their existence, leading to significant challenges in both effectiveness and efficiency of license analysis. To fill this knowledge gap, we conduct a comprehensive empirical study of license variants in the PyPI ecosystem. Our findings show that textual variations in licenses are common, yet only 2\% involve substantive modifications. However, these license variants lead to significant compliance issues, with 10.7\% of their downstream dependencies found to be license-incompatible.

Inspired by our findings, we introduce LV-Parser, a novel approach for efficient license variant analysis leveraging diff-based techniques and large language models, along with LV-Compat, an automated pipeline for detecting license incompatibilities in software dependency networks. Our evaluation demonstrates that LV-Parser achieves an accuracy of 0.936 while reducing computational costs by 30\%, and LV-Compat identifies 5.2 times more incompatible packages than existing methods with a precision of 0.98. 

This work not only provides the first empirical study into license variants in software packaging ecosystem but also equips developers and organizations with practical tools for navigating the complex landscape of open-source licensing.

\end{abstract}

\maketitle

\section{Introduction}
\label{intro}
Open-source software (OSS), with its publicly accessible code and support from package distribution platforms (such as Maven~\cite{Maven}, PyPI~\cite{PyPI}, and npm~\cite{npm}), is now widely reused in modern software development, making third-party OSS integration a common practice~\cite{mohagheghi2007quality,cui2025exploring}. Open-source licenses provide the legal basis for OSS reuse by granting users the rights to use, modify, and redistribute the software~\cite{li2025open,xu2023liresolver}. 
However, these rights come with conditions, as licenses outline specific requirements users must comply with when reusing the code~\cite{Meeker2020OpenSource}. Failure to comply with these conditions can result in license non-compliance and may lead to ethical, legal, and monetary consequences~\cite{liu2024catch, xu2025licoeval}. As OSS packaging ecosystems continue to grow, license incompatibility has become a critical legal risk, which arises when a package (directly or transitively) depends on another package with license containing conflicting conditions~\cite{xu2023understanding}.

As of July 2025, the Open Source Initiative (OSI) has certified 119 licenses~\cite{OSIApproved}, while the SPDX list contains 667 licenses~\cite{SPDX}, ranging from highly restrictive ones (e.g., GPL 3.0) to highly permissive ones (e.g., MIT). For developers, understanding complex legal text is challenging, making it difficult to ensure there are no incompatibilities between licenses during the reuse process~\cite{xu2023licenserec}. The situation becomes even more complicated with the widespread use of license variants~\cite{jahanshahi2025oss},
which are licenses that do not strictly match standard license texts listed by SPDX in either formatting or wording,
or are entirely custom licenses crafted by developers or organizations.

Even small changes in license text can introduce significant differences in terms and conditions.
For instance, as shown in Figure~\ref{fig:usd}, the Python package \Code{usd-core}~\cite{usd-core} (with over 580k downloads on PyPI~\cite{PyPITop} and 1.3k forks on GitHub~\cite{usdgithub}) features a license that differs from the original Apache License 2.0 specifically in Trademarks,
imposing more restrictive conditions, and overlooking this difference may result in non-compliance with trademark usage requirements.
On the other hand, some software owners, particularly commercial entities, choose not to use standard licenses and instead opt for completely custom licenses that better serve their interests~\cite{meloca2018understanding,xu2023liresolver,lindman2010choosing}.
For example, the \Code{cuda-python} package~\cite{cuda-python}, a comprehensive toolkit providing Python access to the NVIDIA CUDA platform, uses a completely customized "\Code{NVIDIA SOFTWARE LICENSE}" that imposes significant restrictions, prohibiting modification and redistribution. Failure to accurately identify these terms could lead to serious consequences.

To help people understand and correctly use the license, previous research has explored license parsing techniques~\cite{xu2023lidetector,gobeille2008fossology,cui2023empirical,cui2025exploring} and compatibility issues in packaging ecosystems~\cite{xu2023understanding,qiu2021empirical,makari2022prevalence}. However, there remain two notable gaps: (1) little is known about the prevalence, characteristics, and impact of license variants; (2) there are no efficient and effective license parsing and incompatibility detection approaches that explicitly account for the presence of these license variants, which still challenge the proper and efficient use of licenses.

\newtcolorbox{licensebox}{colback=gray!10, colframe=black!40, boxrule=0.5pt, arc=3pt, left=4pt, right=4pt, top=4pt, bottom=4pt}

\begin{figure}[tb]
  \centering
  \begin{licensebox}
\footnotesize{
============

OpenUSD

============

Note: The Tomorrow Open Source Technology License 1.0 differs from the
original Apache License 2.0 in the following manner. Section 6 ("Trademarks")
is different.}
  \end{licensebox}
  \caption{Excerpt from the usd-core License}
  \label{fig:usd}
\end{figure}

To bridge the knowledge gaps, we begin with a large-scale empirical study of license variants in the PyPI ecosystem, one of the most active and rapidly growing packaging ecosystems in recent years. We conduct a detailed annotation of license information for the top 8,000 most downloaded packages in the PyPI ecosystem, examining the distribution, impact, and textual characteristics of license variants. 
This analysis is enabled by a comprehensive dataset containing licensing and dependency information for 6,193,233 releases\footnote{In this paper, a release refers to a specific version of a package, aligning with the definition from PyPA~\cite{packaging-glossary}.} from 554,079 PyPI packages.
Our findings reveal that for the vast majority of top PyPI packages, their licenses exhibit minor textual differences from the corresponding standard SPDX license texts. However, only a small fraction (2\%) of these packages involve substantive modifications or fully customized licenses. Despite their small proportion, these substantive license variants have a significant impact: approximately 10.7\% of downstream packages depending on them are found to be license-incompatible. These results show that modifications to license texts are widespread and cannot be ignored, making it risky to simply treat all licenses as their standard versions. Yet, truly substantial changes remain relatively rare. 
Therefore, efficient and accurate license analysis should focus on identifying and interpreting the meaningful differences, rather than exhaustively analyzing every line of license text.

Inspired by our findings, we propose LV-Parser, a license parsing method that explicitly accounts for license variants. By leveraging diff-based analysis, knowledge of standard licenses, and carefully crafted prompts for advanced language models, LV-Parser achieves both high efficiency and accuracy in handling diverse license texts. Building on LV-Parser’s capabilities, we further introduce LV-Compat, a more precise pipeline for detecting license incompatibilities in package dependency networks. Our evaluation demonstrates that LV-Parser achieves high accuracy (0.936) while reducing LLM query costs by nearly 30\% compared to baseline methods. LV-Compat significantly improves incompatibility detection by identifying 5.2 times more incompatible packages than existing methods with 0.98 precision in its detection results.

In summary, the contributions of this paper are as follows:
\begin{itemize}
\item We conduct a comprehensive empirical study on license variants in the PyPI ecosystem, providing key insights into their distribution, impact, and textual characteristics.
\item We propose LV-Parser, an accurate and efficient license parsing method that leverages diff-based analysis and standard license knowledge, offering a novel solution for reliably parsing diverse license texts.
\item We introduce LV-Compat, a precise pipeline for detecting license incompatibilities in package dependency networks, which shows significant improvements in detection accuracy over existing approaches.
\end{itemize}



\section{Background and Related Work}
OSS licensing has been a longstanding focus of research in the software engineering community. In this section, we review prior work in this domain, organizing it into three key areas: license modeling and compatibility, license identification and understanding, and license incompatibility detection.
\subsection{License Modeling and License Incompatibility}
\subsubsection{License Modeling}
\label{sec:Lic-Mod}
OSS licenses are typically written in complex legal language, making them difficult to interpret directly. To address this challenge, prior research has proposed various license modeling approaches that map legal texts to structured representations of key clauses. These models assign labels to relevant license dimensions to support better human understanding and enable automated analysis.
Prior work such as TLDRLegal~\cite{tldrlegal} models OSS licenses using 23 standardized license terms, each representing a type of user action. As shown in Table~\ref{tab: model}, building on this foundation and drawing from earlier approaches~\cite{kapitsaki2019modeling, kapitsaki2017identifying}, the study~\cite{xu2023lidetector} further classified these terms into two categories: Rights and Obligations. Each term is assigned one of three possible values (must, can, or cannot), indicating whether the action is required, permitted, or prohibited under a given license. This structured representation supports clearer license interpretation and compatibility analysis. This classification has been adopted and further refined by many subsequent studies~\cite{xu2023liresolver,li2023lisum,liu2024catch,cui2025exploring,ke2025clausebench,kahol2025oss}. 
\begin{table}[tbp]
\footnotesize
\caption{License Term Categories in previous work.}
\label{tab: model}
\renewcommand{\arraystretch}{1.1}
\setlength{\tabcolsep}{1.5mm}
\begin{tabular}{lp{6cm}}  
    \toprule
    Type & \multicolumn{1}{l}{Term} \\
    \midrule
    Rights & Distribute, Modify, Commercial Use, Relicense, Hold Liable, Use Patent Claims, Sublicense, Statically Link, Private Use, Use Trademark, Place Warranty \\
    \midrule
    Obligations & Include Copyright, Include License, Include Notice, Disclose Source, State Changes, Include Original, Give Credit, Rename, Contact Author, Include Install Instructions, Compensate for Damages, Pay Above Use Threshold \\
    \bottomrule
\end{tabular}
\end{table}
However, this modeling approach has limitations in capturing fine-grained differences across licenses. For example, it cannot effectively distinguish between GPL and LGPL, particularly regarding their different treatments of dynamic linking. It also does not capture the stricter requirements of the AGPL, such as the obligation to disclose source code when software is accessed over a network. Additionally, the model fails to represent other critical legal aspects, such as the termination of patent grants upon litigation, which are present in licenses such as Apache 2.0 and MPL 2.0. These omissions limit the model's ability to support precise and reliable compatibility assessments.

To address these limitations, Xu et al.~\cite{xu2023licenserec,xu2023understanding} propose a more fine-grained license modeling approach. For example, the copyleft dimension (i.e., the requirement to apply the same license to derivative works) is categorized into permissive, file-level, library-level, and strong copyleft types, which is important in the context of package dependencies. Additional dimensions such as network-based source disclosure, termination of patent grants upon litigation, and enhanced attribution requirements are also incorporated. Therefore, in this work, we adopt and refine this license modeling approach. The detailed implementation is described in Section~\ref{LV-Parser}.

\subsubsection{License Incompatibility Definition}
\label{sec:compat-def}
Existing research has provided a clear definition of license incompatibility: \textbf{License A is one-way incompatible with license B if and only if it is infeasible to distribute derivative works of A-licensed software under B}~\cite{kapitsaki2017automating, xu2023licenserec,xu2023liresolver, xu2023lidetector,liu2024catch,xu2023understanding}.
However, there are significant variations in the specific approaches to incompatibility detection. Studies~\cite{xu2023lidetector,xu2023liresolver,cui2025exploring,liu2024catch} suggest that a project's license is considered compatible only if it is the same as or more restrictive than the licenses of its components. This perspective is not well-suited for the context of package dependencies. For instance, a project licensed under MIT (one of the most permissive licenses) can certainly depend on LGPL-licensed third-party packages without requiring the entire project to be under LGPL (one of restrictive licenses), because the LGPL specifically allows such usage patterns~\cite{lgpl-2.1}. Such a definition can result in a high rate of false positives when identifying license incompatibilities.

In the context of package dependencies, Xu et al.~\cite{xu2023licenserec,xu2023understanding} introduce a more precise framework with two types of license compatibility: secondary compatibility and combinative compatibility. The definitions provided by previous works~\cite{xu2023lidetector,xu2023liresolver,cui2025exploring,liu2024catch} correspond to the notion of secondary compatibility, where all parts of a derivative work can be re-licensed under the downstream license. By contrast, combinative compatibility allows for scenarios where a project as a whole may adopt a downstream license, while third-party components retain their original licenses and constraints. For example, an MIT-licensed project depending on an LGPL-licensed package fits the combinative compatibility category, as the LGPL component retains its original terms even when incorporated into the broader project. In this paper, we adopt this definition of compatibility, and the corresponding determination method is described in Section~\ref{LV-Compat}.

\subsection{License Identification and Understanding}

Accurate identification of licenses and license terms is a fundamental step for any license-related research in the OSS domain. Researchers have proposed various automated approaches for license identification from source code, binaries, or license texts~\cite{gobeille2008fossology,tuunanen2009automated,german2010sentence,di2010identifying,liu2019predicting,kapitsaki2017identifying}. 
Recent advances have leveraged neural networks for license term extraction. DIKE~\cite{cui2023empirical}, LiDetector~\cite{xu2023lidetector}, and its successor LiResolver~\cite{xu2023liresolver} all work with the license modeling approach shown in Table~\ref{tab: model}, applying attention networks, Bi-LSTM architectures, and transformer models respectively for increasingly precise license term identification and classification.

More recent work, such as CLAUSEBENCH~\cite{ke2025clausebench}, OSS-LCAF~\cite{kahol2025oss} and LicNexus~\cite{kahol2025oss}, uses large language models (LLMs) for clause recognition and conflict analysis, typically following or refining the license model in Table~\ref{tab: model}. However, these methods generally do not take license variants into account, potentially leading to reduced accuracy when variants are present, or becoming computationally inefficient when scaling to large-scale license analysis.

\subsection{License Incompatibility Detection}

Earlier studies modeled license incompatibility and investigated its manifestation in various systems, including Fedora Linux~\cite{german2010understanding}, Android apps~\cite{duan2017identifying}, and Java applications~\cite{van2014tracing}. 
In packaging ecosystems, Qiu et al.~\cite{qiu2021empirical} and Makari et al.~\cite{makari2022prevalence} reported license incompatibility rates of 0.6\% in npm and up to 13.9\% in RubyGems. Pfeiffer~\cite{pfeiffer2022license} showed that AGPL-related incompatibilities are common across ecosystems, with PyPI and Maven being especially risk-prone.
Xu et al.~\cite{xu2023understanding} provided the first large-scale empirical study of license incompatibilities in PyPI, finding that 7.3\% of releases exhibit incompatibilities, most caused by transitive dependencies, and proposed SILENCE, an SMT-solver-based tool for automated remediation. Other work explored fine-grained, term-level incompatibility detection via argumentation systems~\cite{gordon2011analyzing} and machine learning methods~\cite{xu2023lidetector,cui2023empirical}.

More recently, LLM-based studies including OSS-LCAF~\cite{kahol2025oss}, and LicNexus~\cite{cui2025exploring} have achieved notable progress in clause-level license analysis and conflict detection. However, these approaches generally overlook the existence of license variants and rely on costly full-text parsing, lack precise license modeling which affects detection accuracy, and often use oversimplified compatibility definitions that can lead to high false positive rates in the scenario of package dependencies.

\section{Empirical study on license variants}
\subsection{Research question}
The goal of this empirical study is to provide evidence on the prevalence, characteristics, and impact of license variants in the PyPI ecosystem, which can be the basis of further development of more efficient and accurate license parsing and compatibility detection methods for license variants. To this end, we investigate the following research questions:
\begin{itemize}
    \item \textbf{RQ1:} \textit{What is the prevalence, distribution, and textual characteristics of license variants in the PyPI ecosystem?}
    
    \textbf{Rationale.} Understanding the prevalence and distribution of license variants, as well as their textual deviations from standard licenses, provides crucial empirical insights for designing more effective license analysis methods.
    \item \textbf{RQ2:} \textit{How do license variants impact license compatibility in package dependency networks within PyPI?}
    
    \textbf{Rationale.} This question is motivated by the fact that even minor changes in license terms can lead to significant compatibility issues in complex dependency networks. Analyzing the impact of license variants helps with a more accurate assessment of risks and the development of better compatibility detection methods.
\end{itemize}

\subsection{Study Subject}
We select the top 8,000 most downloaded packages on PyPI as our study subjects. Considering their representativeness and maturity, these packages often exhibit well-established licensing practices~\cite{xu2023understanding}, so that any identified issues are likely a conservative reflection of the broader ecosystem. Given their influence within PyPI, even minor license changes can propagate through dependency networks and potentially lead to significant compatibility issues. 

\subsection{Data Preparation}
\label{sec:datapre}
\subsubsection{Licenses of top packages}
According to~\cite{PyPITop}, we retrieve the latest distribution versions of the 8,000 most downloaded packages from PyPI. We collect both binary distributions and source distributions (if available) and search for license files within the \Code{dist-info} folder. Since including a license file is not mandatory for package releases, we finally obtain 5,990 license files out of the 8,000 packages.

\subsubsection{License and dependency data of PyPI}
\label{sec:data_pre_license}
To assess the impact of packages with license variants on downstream dependencies within the dependency network, we construct a comprehensive dataset containing dependency and license information for the entire PyPI ecosystem as of November 2024. Our method is based on previous work exploring compatibility issues in the PyPI ecosystem ~\cite{xu2023understanding}, which is reported to achieve high accuracy. The specific steps are as follows:

(1) We initiate our analysis by acquiring a comprehensive PyPI distribution metadata dump from the official dataset hosted on Google BigQuery~\cite{PyPIBigQuery} as of November 2024, which consists of 554,079 packages with 6,193,233 distinct releases.

(2) Then we utilize the \Code{requires\_dist} field from the metadata to identify each package’s direct dependencies and their constraints. To obtain the complete dependency tree for each release, including both direct and transitive dependencies, we emulate the breadth-first search behavior of pip while ignoring dependency conflicts and backtracking, finally building the dependency tree for each release in our dataset, with precision and recall achieving 0.9715 and 0.9390, respectively ~\cite{xu2023understanding}. 

(3) Since different versions of a package may adopt different licenses~\cite{vendome2015license}, we collect license data for all releases and extract license information from the \Code{classifier} and \Code{license} fields in the metadata. For releases lacking these fields, we download their distributions and perform license scanning using Scancode~\cite{ScanCode}. Releases without feasible license information are marked as ``unrecognized''. 

Finally, we obtain all the information of the dependency and license for 6,193,233 releases across the entire PyPI ecosystem. 
\begin{figure}
    \centering
    \includegraphics[width=0.85\linewidth]{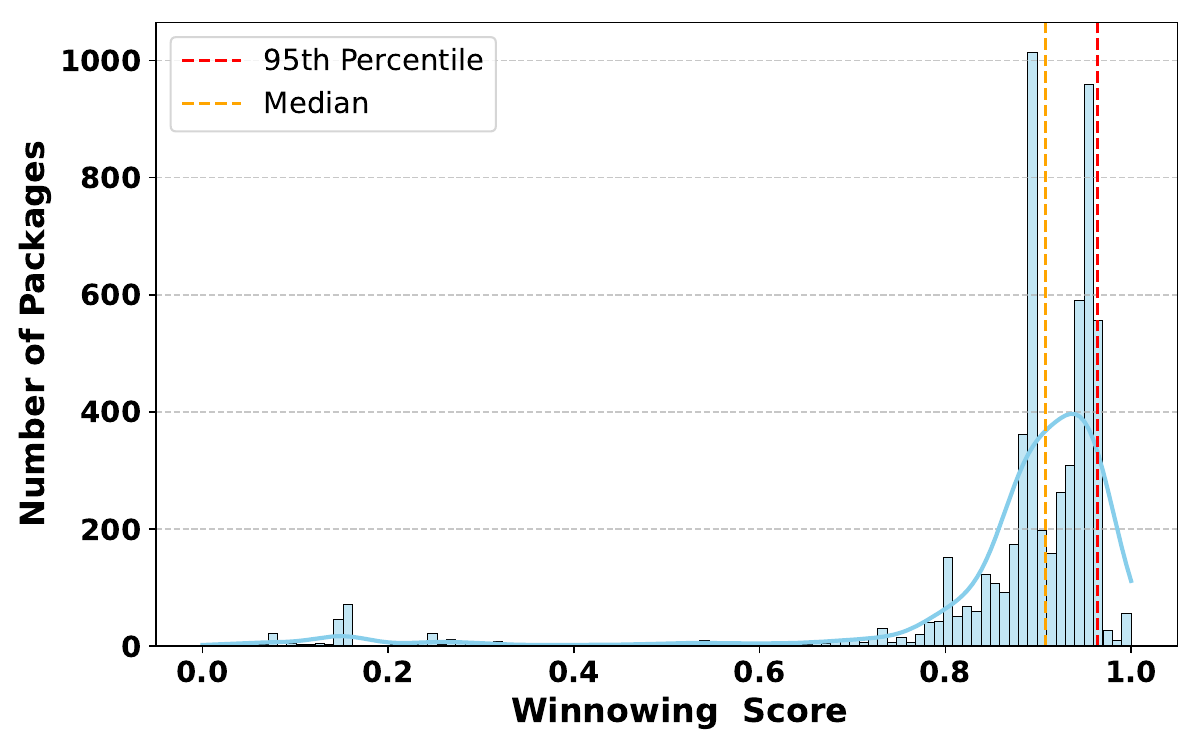}
    \vspace{-3mm}
    \caption{
     Distribution of similarity scores between package licenses and their corresponding standard SPDX licenses
    }
    \vspace{-3mm}
    \label{fig:sim_dis}
\end{figure}
\subsection{Methods and Results}
\subsubsection{RQ1}
We utilize the Winnowing algorithm~\cite{schleimer2003winnowing, jahanshahi2025oss} to compare license files from 5,990 top PyPI packages against standard SPDX license texts. The algorithm creates compact fingerprints by sliding a window over hashed text segments and selecting the smallest hash value per window, enabling efficient and accurate comparisons. 

Following previous work~\cite{jahanshahi2025oss}, for each license file we compute a matching score as the ratio of shared Winnowing signatures to the total signatures between the file and a standard SPDX license, as defined in Equation~\ref{eq:score}:

\begin{equation}
S = \frac{c(A \cap B)}{c(A \cup B)}
\label{eq:score}
\end{equation}

where $S$ is the matching score, $A$ represents the set of signatures in the package license, $B$ represents the set of signatures in the standard SPDX license, and $c(X)$ denotes the count function for the number of elements in set $X$.

The Winnowing similarity scores between each package’s license and its closest matching standard SPDX license text are shown in Figure~\ref{fig:sim_dis}. The distribution shows a median score of 0.90 and a 95th percentile score of 0.96, indicating that textual variations from standard license texts are widespread among package licenses, though most of these differences are minor.

To further understand the nature of these differences, two authors manually review all license texts. Given the complexity of license language, we employ difflib~\cite{difflib}, a line-level diff tool, to avoid overlooking subtle differences during manual inspection. Based on the diff results and the original license texts, the two authors classify the license texts of the top packages into the following four categories:

\begin{itemize}
    \item \textbf{SPDX License}: The license text corresponds to a standard SPDX license, with only minor formatting changes, legally insignificant additions (e.g., \textit{``have fun using this code''}~\cite{burgerbecky}), or is replaced by a license reference (e.g., a name, URL, or a short notice indicating the license type).

    \item \textbf{Dual or Multiple Licenses}: The package is distributed under multiple licenses, so the license file contains the texts of more than one license.
    \item \textbf{Modified SPDX License}: The license is based on a standard SPDX license but includes substantive modifications to its terms and conditions (e.g., the \texttt{usd-core} example mentioned in Section \ref{intro}).
    \item \textbf{Fully Custom License}: The license does not correspond to any standard SPDX license and is entirely custom-written (e.g., the \texttt{NVIDIA} example mentioned in Section~\ref{intro}).
    
\end{itemize}

The two authors independently review and classify all 5,990 license texts along with their corresponding diffs against standard SPDX licenses, with a Cohen’s kappa coefficient of 0.87. In cases where their annotations disagree, a third author is consulted to resolve the disagreement through discussion.
As shown in Table~\ref{tab: distribution}, although textual variations in licenses are common across top PyPI packages, truly substantive modifications that affect the legal terms or conditions remain relatively rare (2\%). These findings highlight the importance of carefully distinguishing between minor, often superficial differences and meaningful changes during license analysis.

\begin{table}

\small
    \caption{Distribution of license texts of TOP packages.}
  \label{tab: distribution}
  \renewcommand{\arraystretch}{1}
  \setlength{\tabcolsep}{1.25mm}
  \vspace{-3mm}
  \begin{tabular}{lrr}
    \toprule
    Type & Count & Percentage \\
    \midrule
    SPDX License&5,869 & 97.98\%  \\
    Dual(or multiple) License& 33 &0.55\%  \\
    Modified SPDX License& 18 & 0.30\% \\
    Custom License& 69 & 1.12\%  \\
    \midrule
    Total &5,990&100\%\\
    \bottomrule
    \end{tabular}
    \vspace{-4mm}
\end{table}
Among the 18 license texts classified as modified SPDX licenses, 11 include exception statements, which represent an important aspect not considered by all previous license modeling approaches~\cite{xu2023lidetector,xu2023licenserec,xu2023understanding}. The remaining 7 involve substantive modifications addressing copyrights, modification clauses, payment requirements, liability disclaimers, feedback obligations, and trademark usage. Notably, one license~\cite{mkl}, although it appears similar to BSD-3-Clause, imposes strict restrictions such as prohibiting modification and the creation of derivative works, making it a proprietary license.

In addition to the four main categories, the authors also identify 144 packages (2.4\%) whose license files include license texts of third-party dependencies. A notable example is the deep learning framework PyTorch, whose license file in the latest binary release (version 2.7.1)~\cite{torch} spans over 8,850 lines and incorporates license texts from hundreds of third-party components. Such complex license texts require additional effort during compliance analysis, as they introduce layered licensing obligations beyond the primary license of the package itself.
\begin{result-rq}{Summary and Implications for RQ1:}
Although license text variations are widespread in the PyPI ecosystem, truly substantive modifications are rare, which can introduce significant compliance concerns, particularly when unexpected restrictions are involved. Furthermore, the inclusion of third-party licenses within a single package adds additional complexity. These results underscore the importance of developing license analysis tools to efficiently distinguish minor textual changes and meaningful legal modifications, and to effectively parse complex, multi-layered license structures.
\end{result-rq}

\begin{table*}
\footnotesize
    \caption{License Model used in LV-Parser.}
  \label{tab: term}
  \renewcommand{\arraystretch}{1}
  \setlength{\tabcolsep}{1.25mm}
  \begin{threeparttable}
  \begin{tabular}{p{2cm}p{8cm}p{7cm}}
    \toprule
    Term & Detailed Definition\tnote{$*$} & Possible Values\tnote{$*$} \\
    \midrule
    Definition\tnote{\dag} & Clause that provides the definition, meaning, or clarification of a specific term, concept, or entity used within the license. & \textbf{0}:Not mentioned or missing definition clause. \textbf{1}:Explicit definition clause exists. \\
    \midrule
    Copyright & Granting rights (or Setting restrictions)  for the software itself (source code, binaries, etc.) on using, copying, modification, distribution, etc. to users or recipients. & \textbf{0\tnote{$*$}} : Proprietary software (prohibits or restricts copying, use, modification, redistribution, or requires payment). \textbf{1}: Public domain (copyright relinquished).  \textbf{2}:Ambiguous copyright grant (implicit or unclear permissions). \textbf{3}: Explicit copyright grant\\
    \midrule
    Copyleft & Requiring any modified versions or derivative works of the software, or works combined with other software, when distributed, to retain the essential terms of the original license or apply a compatible license. & \textbf{0}:Permissive: No restrictions on derivative works; allows relicensing. \textbf{1}:File-level copyleft: Only modified files retain license. \textbf{2}:Library-level copyleft: Only library components retain license. \textbf{3}:Strong copyleft: All derivative/combined works inherit license.\\
    \midrule
    Change Statement& Providing prominent notice marking files, authors, modification dates, or details when distributing the licensed work or its derivatives. & \textbf{0}: No explicit requirement for change notice. \textbf{1}: Explicit requirement to highlight changes. \\
    \midrule
    Patent Grant& Granting users the right to use any relevant patents owned by the original licensors. &  \textbf{-1}: Patent rights denied. \textbf{0}: Patent grant not mentioned. \textbf{1}: Explicitly grants patents. \\
    \midrule
    Trademark Limitation & A clause restricting or prohibiting the use of trademarks related to the licensed software. This typically includes a prohibition on using the names, logos, or trademarks of the software's authors or their organizations for advertising, publicity, or endorsement purposes. &  \textbf{0}: Not mentioned. \textbf{1}: Explicitly restricts or prohibits use of trademark. \\
    \midrule
    Network Use& A clause requiring the disclosure of source code when the licensed work or its derivatives are provided as an online service via network interaction. &  \textbf{0}: Not mentioned. \textbf{1}: Explicit AGPL-style network interaction requirement (source code disclosure, etc.). \\
    \midrule
    Attribution Retention& Requirement to preserve and include the original copyright notice, license text, and/or disclaimer in all copies or distributions of the software, whether in source or binary form, typically within code files or accompanying documents. &  \textbf{0}: No attribution requirement. \textbf{1}: Explicit requirement to retain original attribution/copyright/license. \\
    \midrule
    Enhanced Attribution& A requirement to display or acknowledge the copyright and/or authorship information in a specified or particularly prominent place or form, such as with requirements to display it “prominently,” “in a conspicuous place,” or in designated materials (such as advertising, user interfaces). &  \textbf{0}: No enhanced attribution. \textbf{1}: Explicit requirement for more conspicuous or specific attribution, e.g., specifying where (such as on UI screens, splash screens, product advertising, or other particular places) or how (prominently, with specific wording, format, or frequency) the author or copyright notice must be displayed when distributing. \\
    \midrule
    Secondary License& A clause in open source licenses that permits the distribution, use, or relicensing of the work under a compatible secondary license (for example, work licensed under LGPL-2.1-only may also be relicensed under GPL-2.0-only). & \textbf{None} or \textbf{a list} containing secondary license (e.g. [GPL-3.0-only, GPL-2.0-only]).  \\
    \midrule
    Termination of Patent Litigation& A clause prohibiting users from initiating patent litigation related to the licensed work or its derivatives; violation leads to termination of patent rights granted under the license. &  \textbf{0}: No mentioned. \textbf{1}: Explicit patent termination if user sues. \\
    \midrule
    Termination for 
    Breach\tnote{\dag}&  A clause specifying that the license and associated rights terminate if the licensee breaches the license terms.&  \textbf{0}: No explicit termination-for-breach. \textbf{1}: Explicit license termination clause for breach. \\
    \midrule
    GPL Combination& A clause specifying compatibility and conditions for combining works covered by different GPL licenses or with other software components(e.g., work under LGPL-2.1-only can be combined in work under GPL-2.0-only). &  \textbf{None} or \textbf{a list} of combinatively compatible licenses (only GPL family).  \\
    \midrule
    Compatible Version& A clause indicating whether the licensed work or its derivatives may be re-licensed under later versions of the same license. &  \textbf{None} or \textbf{a list}: Explicitly allows relicense or compatibility with specified license or version (e.g. [GPL-3.0-only, GPL-2.0-only]).  \\
    \midrule
    Explicit Acceptance& A clause requiring that, when distributing the licensed work or its derivatives, the distributor must make reasonable efforts to obtain the recipient's explicit agreement to the license terms. &  \textbf{0}: No mentioned. \textbf{1}: There is a requirement to make reasonable efforts to obtain recipients' explicit acceptance of the license terms when distributing the licensed work or its derivatives.  \\
    \midrule
     Disclaimer\tnote{\dag}& A clause disclaiming warranties or liabilities, stating that the work is provided 'as is', and that the licensors make no representations or warranties with respect to the work. &  \textbf{0}: No warranty disclaimer. \textbf{1}: Explicit no warranty/liability clause.\\
    \midrule
    Governing Law\tnote{\dag}& A clause specifying which country's or region's law applies to the license or to disputes arising under it. & \textbf{None} or \textbf{a list}: Explicitly specifies governing law. \\
    \midrule
    Instruction\tnote{\dag}& A clause describing instructions or guidance on how to use this license. &  \textbf{0}: No explicit instruction/guidance. \textbf{1}: Explicitly gives instructions/guidance.\\
    \midrule
    Usage Limitation\tnote{$*$}& This category covers any explicit clause in the license that restricts or permits only certain types or manners of use of the software. & \textbf{None} or \textbf{a list}: Explicit usage limitation , e.g. [commercial, modify ,SaaS].  \\
    \midrule
    Exception\tnote{$*$}& A clause describing specific exceptions or exemptions to general requirements in the license.& \textbf{None} or \textbf{a list}: Explicit exception to a general rule, e.g. [LLVM, Binary Distribution]. \\
    \midrule
    Other\tnote{\dag}& Any other provisions not covered by the above categories.& -  \\   

    \bottomrule
    \end{tabular}

    \begin{tablenotes}
    \footnotesize
    \item[$*$] Indicates elements that are newly added or clarified in greater detail compared to the modeling methods in~\cite{xu2023licenserec,xu2023understanding}.
    \item[\dag] Terms unrelated to compatibility.
    \end{tablenotes}
    \end{threeparttable}
    
    \label{tab:codemetric}
    
\end{table*}

\subsubsection{RQ2}    
To address RQ2, we investigate the downstream impact of license variants with substantive modifications. From the latter three categories in Table~\ref{tab: distribution}, we select 54 packages that have legally clear license texts, do not involve dual or third-party licenses, and include the complete license text. Using the dependency data that we previously constructed, we identify all downstream packages (both direct and transitive) of the latest release of these 54 packages, resulting in 2,177 downstream packages and 34,004 releases.


As described in Section~\ref{sec:compat-def}, the definition of secondary and combinative license compatibility fits well within the context of packaging ecosystems~\cite{xu2023understanding}, where a package may be considered a derivative work of its dependencies, according to the interpretation of the Free Software Foundation (FSF)~\cite{WhatIsDeriveWork}. Based on this definition and the compatibility check methodology proposed by Xu et al.~\cite{xu2023licenserec}, we manually review and annotate the 54 license texts along key legal dimensions, including copyright grant, copyleft level, patent grant, trademark grant, usage restrictions, etc. Note that our goal is to make a conservative estimate of incompatibilities within the PyPI ecosystem, so it is acceptable that the dimension set may not be strictly exhaustive. Then, we adopt widely accepted principles derived from previous work and official interpretations~\cite{xu2023understanding,xu2023licenserec,gnu_gpl_faq,apache_faq,cecill_faq,gnu_license_list,eclipse_faq,eupl_guidelines} (e.g., FSF explicitly states that free software cannot depend on proprietary software due to fundamental license incompatibility~\cite{gnu_gpl_faq}) to identify license incompatibilities of the 54 packages and their downstream releases. Any cases in which compatibility could not be determined with confidence are conservatively labeled as \Code{unknown}.

Our analysis reveals that 234 (10.7\%) out of the 2,177 downstream packages exhibit license incompatibilities with their upstream variants. At the release level, 3,070 (9.0\%) out of 34,004 releases are found to be incompatible, while 7,426 releases cannot be conclusively assessed and are labeled as ``unknown''. Among the 3,070 incompatible releases, 1,879 cases (5.5\% of the total) involve license variants that impose proprietary prohibitions on copyright-related rights such as use, modification, or distribution. In these cases, downstream packages are released under free software licenses while depending on upstream proprietary software---explicitly deemed incompatible under FSF guidance~\cite{gnu_gpl_faq}.

While standard SPDX licenses can themselves lead to downstream incompatibilities, modifications in variants greatly amplify this complexity. In our analysis of licenses altered relative to their SPDX originals, we identify two contrasting patterns: some variants introduce stricter obligations, leading to a 21\% increase in incompatible downstream releases compared with their SPDX counterparts, whereas others add exceptions that relax obligations, resulting in a 76\% reduction in incompatibilities. These findings indicate that license variants can exert opposite effects on compatibility and underscore the need for dedicated analysis methods that can effectively capture such modifications to improve recall while reducing false positives. The 7,426 uncertain releases (21.8\% of the total) further highlight the substantial challenges and the limitations of current compatibility frameworks when confronted with non-standard or modified license texts.

\vspace{-1mm}
\begin{result-rq}{Summary and Implications for RQ2:}
Our analysis reveals significant compatibility issues caused by license variants, with 10.7\% downstream packages affected. Most critically, we find 1,879 cases in which variants lead to incompatibilities with downstream free software packages due to the proprietary prohibitions they have introduced. These findings highlight the importance of carefully examining license variants for clauses that restrict use, modification, or distribution rights, as even subtle alterations can effectively transform an open-source license into a proprietary one.
\end{result-rq}

\section{The LV-Parser approach}
\label{LV-Parser}
Inspired by the empirical study, we propose LV-Parser, a license parsing approach that explicitly accounts for license variants. In this section, we describe our license modeling methodology, the design of LV-Parser, and its evaluation.
\subsection{License Modeling}
\label{sec:licmod}
License modeling serves as the foundation for understanding licenses and analyzing compatibility. As discussed in Section~\ref{sec:Lic-Mod}, existing modeling approach~\cite{xu2023lidetector,tldrlegal,xu2023liresolver} in Table~\ref{tab: model} demonstrates significant limitations that result in incomplete license representations. To address these shortcomings, we adopt the modeling framework proposed in~\cite{xu2023licenserec,xu2023understanding} and further refine it based on insights from our empirical study. As shown in Table \ref{tab: term}, it includes both compatibility-related and unrelated terms, providing a complete overview of the terminology used in our license modeling. 

It is worth noting that our empirical analysis found that substantive license variants frequently include exception clauses and introduce usage restrictions, which have not been adequately addressed by previous work~\cite{xu2023licenserec, xu2023understanding}. Thus, we expand our model to include two new dimensions: usage restrictions and exceptions. In addition, we enhance the copyright dimension by introducing a proprietary software category, enabling our model to more accurately capture licenses that, despite their appearance, impose restrictive terms comparable to proprietary software. Based on this enhanced model, we develop LV-Parser, a method designed to accurately interpret and analyze software licenses.
\begin{figure*}[t]
    \centering
    \includegraphics[width=0.95\linewidth]{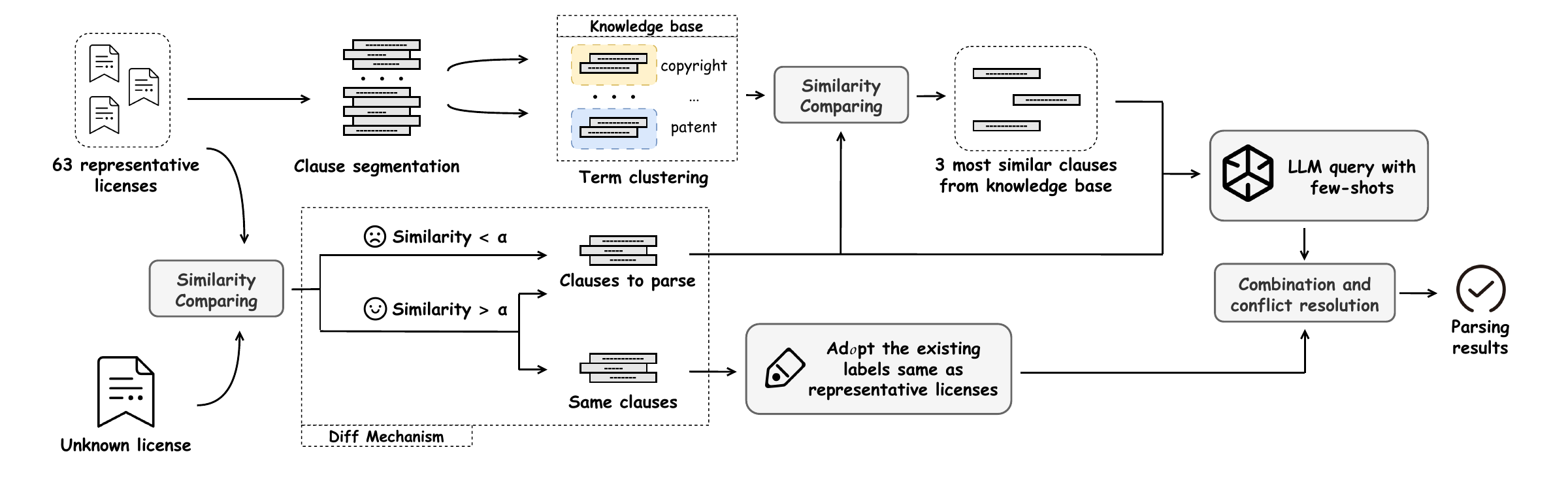}
    \caption{
     Overview of license parsing methodology
    }
    \label{fig:parse}
\end{figure*}

\subsection{Design of LV-Parser}
Inspired by our empirical findings that only a small fraction of license variants introduce substantive changes to legal terms despite their prevalence, we design a diff-based analysis approach for LV-Parser, as shown in Figure~\ref{fig:parse}. Our method avoids redundant analysis of standard provisions by leveraging established knowledge of license-related terms. And we utilize LLMs with advanced text understanding capabilities to assist in interpreting and analyzing the nuanced differences present in variant license texts.
\subsubsection{Fine-grained Knowledge base construction}

Since we have revealed that widespread license variants only lead to minor substantial differences, it is unnecessary to repeatedly analyze identical portions when reviewing specific licenses. Instead, we construct a fine-grained knowledge base by thoroughly annotating representative standard licenses, which enables us to reuse these annotations for identical sections and focus our analysis only on the different segments when examining each variant. Additionally, this knowledge base can provide valuable context to help LLMs analyze the textual differences. We detail the steps for constructing the knowledge base as follows. 

First, we select 63 representative licenses, focusing on those that (1) are certified by the FSF or OSI, (2) are current and not obsolete (e.g., excluding Apache-1.1), and (3) are not restricted to specific domains, software, or authors (such as the IPA font license)~\cite{xu2023licenserec,xu2023understanding}. This set covers 95.2\% (5,703) of the 5,990 license texts identified in our empirical study, providing both high coverage and diverse representation of the ecosystem, so that it has the potential to support efficient and accurate analysis of most licenses in practice.
Then, to build the knowledge base, we segment each license into individual sentences, resulting in 3,510 clauses, with each representing a self-contained unit of meaning. Two authors independently classify these sentences according to the license term types defined in Table~\ref{tab: term}. Since some sentences may pertain to more than one term (e.g., addressing both patent grants and patent litigation), we employ a multi-label classification process. Sentences that do not fit any existing term are labeled as \textit{``other''}. The initial annotation achieves a Cohen’s kappa of 0.84, and all discrepancies are resolved through discussion with a third author.

For each license, we then aggregate all sentences associated with the same term to produce a comprehensive set of clauses for each dimension. By leveraging our detailed annotation together with existing labeled data from~\cite{xu2023licenserec,xu2023understanding}, we determine the specific values for every term in each license. Ultimately, our knowledge base contains 691 annotated terms across the 63 representative licenses. To facilitate efficient matching during parsing, we use inf-retriever-v1~\cite{infly-ai_2025}, an LLM-based dense retrieval model, to compute embeddings for each license and each term in our knowledge base. We select this model because it is the best embedding model in the legal domain on the AIR-Bench benchmark~\cite{yang2024air} as of January, 2025. These embeddings allow us to quickly identify similar or relevant text segments when analyzing new license variants. 
\subsubsection{Parsing Method using LLMs}

To avoid redundant analysis while focusing on substantive differences, LV-Parser adopts a diff-based approach. The overall workflow is illustrated in Figure~\ref{fig:parse} and proceeds as follows:

First, we use INF-Retriever-v1 to compute an embedding for the entire license text to be analyzed. We then compare this embedding with those of the 63 standard licenses in our knowledge base. If the similarity with the most similar standard license exceeds a pre-defined threshold (set to 0.9 in our experiment), we consider the license under analysis to be a variant of that standard license.

Next, we segment both the analyzed license and the matched standard license into sentences. For sentences in the analyzed license that are exactly the same as those in the standard license, we directly adopt the corresponding term values from the knowledge base. Sentences that do not have an exact match are flagged for further analysis. If the overall license similarity falls below the threshold, we treat the license as not matching any standard license and proceed to analyze all sentences individually.

For all sentences requiring further examination, we perform multi-label classification using GPT-4.1~\cite{GPT-4.1}, a state-of-the-art LLM, with prompts that incorporate the detailed term definitions from Table~\ref{tab: term}. Considering a sentence may pertain to multiple terms, we adopt a deliberately greedy classification strategy: the prompt explicitly instructs the language model to include a sentence under a term whenever there is any uncertainty about its relevance to that category. This approach errs on the side of inclusion, ensuring that no potentially relevant text is missed.

This design choice is motivated by the fact that over-inclusion of sentences in a given category does not negatively affect the subsequent determination of term values. For example, if a copyright-related sentence is incorrectly included under "patent," the language model will still infer the patent term’s value based primarily on sentences that are truly relevant. In contrast, missing a sentence that should be included (i.e., under-inclusion) may result in a critical clause being left out of a term’s analysis and thus lead to incorrect term assignment.

After the initial classification, we aggregate all sentences under each term. For each term, we retrieve the top three most similar examples from our knowledge base using dense retrieval (based on the pre-computed embeddings), along with their assigned values. We then prompt the language model to determine the most appropriate value for the analyzed term, taking into account the detailed definition from Table~\ref{tab: term} and the top three similar examples.

If any conflicts arise between automatically assigned values (from identical sections) and values inferred from non-matching sections, we conservatively adopt the more restrictive value to ensure compliance and reduce the risk of misclassification.

In summary, LV-Parser combines embedding-based matching, sentence-level diffing, knowledge reuse, and LLM-assisted interpretation to efficiently and accurately parse license variants, while minimizing redundant analysis and focusing effort on substantive textual differences.

\begin{table*}
\footnotesize
    \caption{Accuracy of different license parsing approaches.}
  \label{tab:parser-res}
  \renewcommand{\arraystretch}{1}
  \setlength{\tabcolsep}{0.5mm}
  \begin{tabular}{lcccccccccccccccc}
    \toprule
     \makecell{\rotatebox{45}{Approach}} & \makecell{\rotatebox{45}{copyright}} & \makecell{\rotatebox{45}{copyleft}} & \makecell{\rotatebox{45}{change\_state}} & \makecell{\rotatebox{45}{patent}}&\makecell{\rotatebox{45}{trademark}}&\makecell{\rotatebox{45}{network\_use}}&\makecell{\rotatebox{45}{retain\_attr}}&\makecell{\rotatebox{45}{enhance\_attr}} & \makecell{\rotatebox{45}{patent\_term}} & \makecell{\rotatebox{45}{acceptance}} & \makecell{\rotatebox{45}{secondary}} & \makecell{\rotatebox{45}{compat\_ver }}& \makecell{\rotatebox{45}{gpl\_combine}} & \makecell{\rotatebox{45}{usage\_limit}} & \makecell{\rotatebox{45}{exception}} &  \makecell{\rotatebox{45}{Mean}}\\
    \midrule
    LV-Parser&\textbf{0.757} & 0.878& 0.973&0.946& \textbf{1.0}& \textbf{1.0}& 0.973 & \textbf{0.973} &\textbf{1.0} & \textbf{0.973} & \textbf{1.0/2} & \textbf{3.3/4} & \textbf{3.8/4}& \textbf{29.8/38} & 3.5/5 &\textbf{0.936} \\
    Baseline& 0.635 &0.892 & 0.959& 0.959&0.986&1.0&0.905&0.892&0.973&0.905 &0.25/2 &1.47/4&  0/4&29.8/38 &3.5/5 &0.894  \\
    \midrule
    Claude-4-sonnet& 0.514 & 0.919 & 0.986 & 0.892 &0.946 & 1.0 &0.932 & 0.959 & 1.0 & 0.973 & 1.0/2 &3.27/4 &3.8/4 & 28.3/38 & 3.0/5 &0.901 \\
    Deepseek-v3& 0.554 & 0.838 &0.919&0.919 &0.932 & 0.986 &0.986 &0.959 &0.973 &0.919 &1.0/2 &3.27/4 &3.8/4 &26.1/38&4.5/5 &0.887  \\
    Qwen3-30B & 0.581 &0.865 &0.959 & 0.865 &0.878 & 1.0 & 0.973 &0.959&0.946&0.838&1.0/2 &3.27/4 &3.8/4 &17.9/38&1.0/5&0.861\\
    Qwen3-14B& 0.595 & 0.973 &0.986&0.919&0.946& 1.0 &  0.973 & 0.973 & 1.0 &0.919 &1.0/2 &3.27/4 & 3.8/4 & 13.3/38 &1.0/5 &0.895 \\
    
    \bottomrule
    \end{tabular}
\end{table*}
\subsection{Evaluation of LV-Parser}
\label{sec:LVP-eve}
\subsubsection{Evaluation Tasks and Dataset}

There are two key steps in LV-Parser that utilize LLMs for classification: categorizing sentences into license terms and determining the value for each term. However, since we deliberately employ a greedy approach in the first step, explicitly allowing the LLM to assign a sentence to as many categories as possible, we do not evaluate the accuracy of this step independently. Instead, we adopt an end-to-end evaluation that focuses solely on the final parsing results produced by the model.

Our evaluation dataset consists of the following samples:
\begin{itemize}
    \item The 54 licenses from RQ2 in our empirical study, which are either substantially modified versions of standard licenses or fully custom licenses, with complete license texts and without complications such as third-party licenses or dual licensing.
    \item For the first category in Table~\ref{tab: distribution} of our empirical study (projects using standard SPDX licenses), there are 83 SPDX licenses involved in total. Since 63 of them are already included in our knowledge base, we randomly select one sample from projects using each of the remaining 20 SPDX licenses, which may contain minor textual differences but no substantive changes, for evaluation.
\end{itemize}

In total, our evaluation dataset comprises 74 licenses: 20 standard SPDX licenses with only minor textual differences from the reference texts and 54 licenses with substantial modifications or custom terms.
The two authors independently annotate all 74 licenses in the evaluation dataset set according to each term defined in Table~\ref{tab: term}. The inter-annotator agreement, measured by Cohen’s kappa, is 0.86. Any disagreements are resolved through discussion with a third author. The final annotations serve as the ground truth for our evaluation.
\subsubsection{Experiment Setup}

Recent research has utilized LLMs for license parsing and showed promising results~\cite{cui2025exploring,ke2025clausebench,kahol2025oss}. However, these works adopt the license modeling framework from ~\cite{xu2023lidetector} as shown in Table~\ref{tab: model}, which differs from our approach and makes it difficult to directly compare with LV-Parser. To provide a fair baseline, we adopt their core methodology as the baseline implementation: the baseline model is provided with rich context, including term definitions and value definitions, along with the target license text, and is asked to determine the appropriate value for each term.
To ensure a fair comparison, the baseline uses the same term and value definitions as LV-Parser and employs the same LLM, GPT-4.1. The only difference is that the baseline does not incorporate diff-based processing or provide similar example clauses as additional context. In all experiments, the LLM operates in greedy decoding mode with a temperature setting of zero.
Additionally, we evaluate our approach by replacing GPT-4.1 with several different LLMs (including Claude-4-Sonnet~\cite{claude}, DeepSeek-v3~\cite{liu2024deepseek}, Qwen3~\cite{yang2025qwen3}) to demonstrate the robustness and generalizability of our parsing methodology, regardless of the specific model used.
\subsubsection{Results}
As shown in Table~\ref{tab:parser-res}, the first ten terms have numerical (single-value) labels, and the table presents the accuracy of the model on all 74 samples for these terms. The last five terms are open-ended (multi-value) terms, each with a list of possible values. For these, the table reports the average recall across samples that include each term. Specifically, the denominator is the number of relevant samples, and for each sample, recall is calculated as the intersection between the ground truth and predicted lists divided by the length of the ground truth list.

The experimental results show that LV-Parser achieves strong performance in license parsing across most dimensions, with an average accuracy of 0.936, outperforming the baseline’s 0.894. In terms of efficiency, we compare the number of LLM queries required by each method. On average, LV-Parser calls the LLM 42 times per license, compared to 59 times for the baseline. Even with our relatively conservative similarity threshold ($\alpha$ = 0.9, below which full-text analysis is required), LV-Parser still reduces the number of queries by nearly 30\%. This substantial reduction in LLM invocations suggests significant improvements in computational efficiency and cost-effectiveness. Overall, our results demonstrate that LV-Parser not only maintains superior accuracy but also offers greater efficiency in license parsing.

Additionally, when replacing GPT-4.1 with other LLMs in LV-Parser, we observe only a minor drop in performance. Remarkably, even when using the smaller and less powerful Qwen3-14B model, LV-Parser attains an average accuracy of 0.895. This result highlights the strong robustness of our diff-based method and knowledge-enhanced prompting: even models with limited capacity can achieve high-quality results when equipped with effective context and relevant examples provided by our approach. This demonstrates  that our method's effectiveness stems primarily from its well-structured analysis framework rather than relying on the capabilities of any specific language model.

\begin{figure*}[t]
    \centering
    \includegraphics[width=0.99\linewidth]{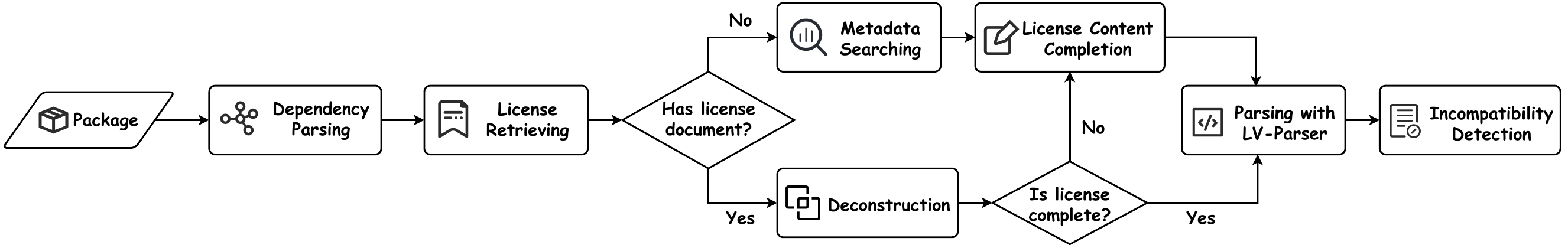}
    \caption{
     Overview of license incompatibility detection pipeline
    }
    \label{fig:compa}
\end{figure*}

\section{The LV-Compat Pipeline}
\label{LV-Compat}
\subsection{Incompatibility Determination Method}
Based on our license modeling approach described in Section~\ref{sec:licmod}, we propose two algorithms for checking license compatibility that align with the more nuanced framework introduced by Xu et al.~\cite{xu2023licenserec,xu2023understanding}. This framework distinguishes between secondary compatibility (where derivative works can be entirely re-licensed under the downstream license) and combinative compatibility (where components retain their original licenses while the project adopts a different license).

Our algorithms, presented in Algorithm~\ref{alg:cc}, center on copyleft terms as the primary determinants of compatibility in package dependencies. The compatibility check begins by extracting core terms from both licenses, with special focus on their copyleft levels. For exact license matches, both compatibility types are automatically satisfied. In other cases, the algorithm evaluates whether the downstream license can satisfy the upstream license's conditions, considering secondary compatibility when all terms can be satisfied, and combinative compatibility for scenarios such as MIT-licensed projects depending on LGPL components.

The implementation additionally incorporates specialized treatment for exception clauses that may alter standard compatibility relationships, a significant enhancement we introduce over~\cite{xu2023licenserec,xu2023understanding} to better reflect real-world licensing complexities. 
\subsection{Design of LV-Compat}

Software package license texts often present significant complexity, frequently encompassing multiple overlapping licenses, third-party terms, and instances of missing or incomplete information, as revealed by our empirical study. To tackle these challenges, we propose LV-Compat, an automated pipeline for large-scale compatibility checking, built atop LV-Parser.

As shown in Fig.\ref{fig:compa}, the pipeline initiates with comprehensive dependency resolution, systematically identifying both direct and transitive upstream dependencies by employing the methodology described in Section~\ref{sec:datapre}. Upon obtaining each software distribution, the pipeline performs license extraction at scale. Recognizing the prevalence of multi-license and composite license files, LLMs (GPT-4.1) are leveraged to accurately segment and structurally disambiguate license texts into discrete, machine-readable components. The same models assess license completeness, automatically augmenting incomplete entries with canonical SPDX license texts if SPDX identifiers are detected. In scenarios lacking explicit license text, relevant metadata fields are mined to fill informational gaps.

Once license content is distilled into a structured representation, LV-Parser enables fine-grained term-level analysis for each identified clause or provision. Subsequently, a compatibility reasoning module evaluates all third-party dependencies, applying compatibility determination algorithms to assess legal interoperability across the entire dependency graph. This end-to-end analytical pipeline is underpinned by the comprehensive dataset constructed during empirical study described in Section~\ref{sec:datapre}, ensuring high-fidelity compatibility assessment within complex package ecosystems.

\begin{algorithm}[bt]
\footnotesize
\caption{License Compatibility Check}
\label{alg:cc}
\DontPrintSemicolon
\SetKwInput{Input}{Input}
\SetKwInput{Output}{Output}
\Input{Two licenses $L_1$, $L_2$ with terms $T_1$, $T_2$}
\Output{Compatibility $\mathcal{C} \subseteq \{\mathbf{S}, \mathbf{C}, \mathbf{I}\}$ \\
\tcc{$\mathbf{S}$: Secondary, $\mathbf{C}$: Combinative, $\mathbf{I}$: Incompatible}
}
$\mathcal{C} \leftarrow \emptyset$ \;
Extract copyleft levels $c_1$, $c_2$, term sets $t_1$, $t_2$, and compatibility lists $\mathcal{V}$, $\mathcal{S}$, $\mathcal{G}$ from $T_1$, $T_2$\;

\tcc{$t$: all terms except copyleft, $\mathcal{V}$: version-compatible list, $\mathcal{S}$: secondary-compatible list, $\mathcal{G}$: GPL-combinable list}

\If{$\text{IsSecondaryCompatible}(T_1, T_2)$}{
    $\mathcal{C} \leftarrow \mathcal{C} \cup \{\mathbf{S}\}$ \;
}
\If{$\text{IsCombinativeCompatible}(T_1, T_2)$}{
    $\mathcal{C} \leftarrow \mathcal{C} \cup \{\mathbf{C}\}$ \;
}
\If{$\text{HasSpecialException}(T_1, T_2)$}{
    $\mathcal{C} \leftarrow \text{ApplyExceptionRules}(\mathcal{C}, T_1, T_2)$ \;
}
\If{$\mathcal{C} = \emptyset$}{
    $\mathcal{C} \leftarrow \{\mathbf{I}\}$ \;
}
\Return $\mathcal{C}$\;

\medskip
\SetKwFunction{FSecondary}{IsSecondaryCompatible}
\SetKwProg{Fn}{Function}{:}{}
\Fn{\FSecondary{$T_1$, $T_2$}}{
    Extract $c_1$, $c_2$, $t_1$, $t_2$, $\mathcal{V}$, $\mathcal{S}$ from $T_1$, $T_2$
    \uIf{$c_1 = 0$ \textbf{and} ($t_1 \subseteq t_2$ \textbf{or} $L_2 \in \mathcal{V}$ \textbf{or} $L_2 \in \mathcal{S}$)}{
        \Return true\;
    }
    \uElseIf{$c_1 > 0$ \textbf{and} $c_2 \geq c_1$ \textbf{and} ($t_1 \subseteq t_2$ \textbf{or} $L_2 \in \mathcal{V}$ \textbf{or} $L_2 \in \mathcal{S}$)}{
        \Return true\;
    }
    \Return false\;
}
\SetKwFunction{FCombinative}{IsCombinativeCompatible}
\Fn{\FCombinative{$T_1$, $T_2$}}{
    Extract $c_1$, $c_2$, $\mathcal{G}$ from $T_1$
    \If{ ($c_1 = 0$ \textbf{and} $c_2 \in \{0,1\}$) \textbf{or}
         ($c_1 \in \{1,2\}$ \textbf{and} $c_2 \in \{0,1\}$) \textbf{or}
         ($c_1 > 0$ \textbf{and} $L_2 \in \mathcal{G}$)
    }{
        \Return true\;
    }
    \Return false\;
}
\end{algorithm}

\subsection{Evaluation of LV-Compat}
To evaluate the effectiveness of LV-Compat in assessing the compliance of real-world software packages, we use the same 74 packages as described in Section~\ref{sec:LVP-eve}. From all 3,449 of their downstream releases, we randomly sample 75 releases to serve as the evaluation set for LV-Compat. We compare LV-Compat with SILENCE~\cite{xu2023understanding}, which utilizes the same license modeling approach and license definitions. As described in Section~\ref{sec:data_pre_license}, SILENCE obtains license information without considering variants. Furthermore, its compatibility analysis is limited to the predefined set of 63 licenses, meaning that any licenses outside this set cannot be assessed for compatibility.
The results, shown in Table~\ref{tab:res-LVC}, highlight that LV-Compat substantially outperforms SILENCE in real-world compliance assessment. Among the 75 downstream packages evaluated, each package was checked for compatibility with all its dependencies. ``Package'' columns in Table~\ref{tab:res-LVC} reflect cases where packages were found to have incompatible, unknown, or compatible dependencies.

LV-Compat identified 52 packages with at least one incompatible dependency, compared to just 10 for SILENCE. In addition, SILENCE was unable to determine compatibility for 65 packages, whereas LV-Compat reduced the number of packages marked as ``Unknown'' to only 7.
On the dependency level, the ``Dependency'' columns in the table record the total number of dependency relationships found incompatible or unknown. LV-Compat detected 116 incompatible dependencies, far more than the 30 identified by SILENCE, and decreased the number of dependencies marked "Unknown" from 598 to 52.
A key advantage of LV-Compat is its ability to parse and analyze third-party software licenses embedded within dependency license text, a dimension SILENCE does not cover. This further analysis uncovered 238 incompatibilities and 148 unknowns relating to third-party components. By incorporating third-party software considerations, LV-Compat delivers a far more comprehensive assessment of license compatibility and compliance compared to prior approaches.

To evaluate the accuracy of these results, we conduct a targeted validation. Since verifying that a package has no incompatibilities would require checking all its dependencies' licenses, and given the inherent ambiguities in the legal domain that make ground truth difficult to establish, we instead randomly select one dependency from each of the 52 packages that LV-Compat reported as having incompatible dependencies. Upon examining the original license texts, we confirmed that 51 of these dependencies were indeed incompatible, with only one error stemming from LV-Parser incorrectly labeling a package's copyright dimension. This translates to a precison of 0.98, demonstrating that LV-Compat achieves its substantially higher incompatibility detection rate without introducing a significant number of false positives.

\begin{table}[t]

\footnotesize
    \caption{Results of LV-Compat and SILENCE.}
  \label{tab:res-LVC}
  \renewcommand{\arraystretch}{1}
  \setlength{\tabcolsep}{0.5mm}
  \begin{tabular}{l|ccc|cc|cc}
    \toprule
    \multirow{2}{*}{Approach} & \multicolumn{3}{c|}{Package}  &\multicolumn{2}{c|}{Dependency} &\multicolumn{2}{c}{3rd-party software} \\
    &Incom. &Unknown &Compat. &Incom. &Unknown &Incom.&Unknown\\
    \midrule
    LV-Compat& 52 & 7 & 16 &116 &52 &238 &148  \\
    SILENCE& 10 & 65 & 0 & 30 & 598& - &-\\

    \bottomrule
    \end{tabular}
\end{table}

\section{Discussion}
\subsection{Case study}
We examine terms where LV-Parser's performance is less satisfactory, such as the copyright dimension. Since this is a four-class classification task, it is inherently more challenging. Most misclassifications occurred between category 2 (ambiguous copyright grant) and category 3 (explicit copyright grant), as the boundary between the two is often unclear: the former involves cases where the author does not make a clear statement, but implicitly grants rights related to use, modification, and distribution. For example, in the package \Code{lmdb}~\cite{lmdb}, the license statement "redistribution and use of this software and associated documentation ('software'), with or without modification, are permitted provided that..." is manually annotated as an ambiguous copyright grant, while the LLM classify it as an explicit grant. However, these misclassifications have no impact on overall compatibility results.

Another challenging term is usage limitation, which is modeled as an open-ended list of restrictions. Due to its open nature, this term remains difficult to classify precisely. Addressing these open-list value terms presents a promising direction for future improvements.

There are also examples where the complexity of legal language or the need for deeper legal expertise leads to disagreements even among authors during annotation. For instance, in the license of the package \Code{encodec}~\cite{encodec}, the clause "Patent and trademark rights are not licensed under this Public License." was annotated by human annotators as "not mentioned" for the patent grant term, while LV-Parser classified it as "explicitly not granted." According to Meeker~\cite{Meeker2020OpenSource}, even in the absence of an explicit patent clause, most licenses are considered to implicitly grant some patent rights, although there is little relevant case law. Therefore, after discussion, the annotators chose to mark the clause as 0 ("not mentioned") rather than -1 ("explicitly not granted").

Furthermore, we also manually examine the six packages for which LV-Compat is unable to determine compatibility (labeled as "unknown"). In all cases, the root cause is incomplete license information for one or more upstream dependencies. For example, some dependencies specified only "BSD" without indicating the license version, which made it infeasible to accurately assess compatibility.
\subsection{Limitation}
In this section, we discuss several notable limitations of the empirical study, LV-Parser, and LV-Compat. 

For the empirical study, our analysis primarily targets the most popular PyPI packages, which may not reflect the full diversity of the ecosystem. The compatibility assessment is based on methods from prior work, which may introduce some inaccuracies; however, it provides a conservative and practical estimate of licensing risks. Both the LV-Parser knowledge base and evaluation datasets are manually annotated, which introduces potential subjectivity and may limit consistency, especially given the complexity and ambiguity of legal language.
Finally, while LV-Compat achieves high precision in incompatibility detection, its recall cannot be fully assessed due to the impracticality of exhaustively checking every dependency and the inherent uncertainty in legal interpretation. Despite these limitations, the substantial increase in detected incompatibilities compared to baseline methods demonstrates improved effectiveness. Additionally, this study primarily focuses on license incompatibility arising from package dependencies and does not take into account license information embedded in source code, which may not be included in the package distribution. However, our LV-Parser and LV-Compat are highly extensible and can be readily applied to this scenario, making it a valuable direction for future work.

In terms of external validity, our findings and approaches may not generalize completely to other software ecosystems beyond PyPI, as different ecosystems may have distinct licensing practices, variant patterns, and dependency structures. Future research could validate and adapt our techniques across multiple packaging platforms to strengthen the generalizability of our results.
\section{Conclusion}
In this paper, we conduct the first comprehensive empirical study on license variants in the PyPI ecosystem, analyzing their distribution, impact, and textual characteristics. 
Inspired by the findings, we develop LV-Parser, an efficient license parsing method that leverages diff-based analysis, knowledge of standard licenses, and LLMs to accurately interpret license variants. Building on LV-Parser, we create LV-Compat, a comprehensive pipeline for detecting license incompatibilities in package dependency networks. Our evaluation demonstrates that LV-Parser achieves high accuracy (0.936) while reducing computational costs by nearly 30\%, and LV-Compat identifies 5.2 times more incompatible packages than existing methods with 0.98 precision.

This work makes significant contributions to understanding and addressing license compliance challenges in modern software ecosystems. By providing tools that can effectively parse license variants and detect incompatibilities, we help developers and organizations navigate the complex landscape of open-source licensing. As software development continues to rely heavily on third-party dependencies, approaches like LV-Parser and LV-Compat will become increasingly valuable for ensuring legal compliance and reducing risk in software supply chains. 
\section{Data Availability}
We make our dataset and code available at \url{https://figshare.com/s/4fbaedbeb120d1940f12}.

\begin{acks}
This work is sponsored by the National Natural Science Foundation of China 62332001, 62502030, and Fundamental Research Funds for the Central Universities FRF-TP-25-030.
\end{acks}

\bibliographystyle{ACM-Reference-Format}
\bibliography{main}


\end{document}